\documentclass[conference]{IEEEtran}
\IEEEoverridecommandlockouts

\usepackage{cite}
\usepackage{amsmath,amssymb,amsfonts}
\usepackage{algorithmic}
\usepackage{graphicx}
\usepackage{textcomp}
\usepackage{xcolor}
\def\BibTeX{{\rm B\kern-.05em{\sc i\kern-.025em b}\kern-.08em
    T\kern-.1667em\lower.7ex\hbox{E}\kern-.125emX}}

\usepackage{inconsolata}
\usepackage{booktabs}
\usepackage{multicol,multirow}
\usepackage{makecell}
\usepackage{hyperref}
\hypersetup{hidelinks,
	colorlinks=true,
	allcolors=black,
	pdfstartview=Fit,
	breaklinks=true}
\usepackage{setspace}
\usepackage{amssymb}

\usepackage[ruled,linesnumbered, ,vlined,noresetcount]{algorithm2e}

\usepackage{bm}
\usepackage{color}
\usepackage{tikz}
\newcommand*\circled[1]{\tikz[baseline=(char.base)]{\node[shape=circle,draw,inner sep=0.6pt] (char) {\footnotesize{#1}};}}
\usepackage{pifont}
\usepackage{marvosym}

\usepackage{draftwatermark}
\SetWatermarkText{Accepted by ICCD 2026}
\SetWatermarkScale{0.45} 
\SetWatermarkColor[gray]{0.90} 

\begin{document}

\title{CMD: An Integrated CGRA Framework with Cluster-Based Distributed Memory Design}

\author{
\IEEEauthorblockN{
Shangkun Li\IEEEauthorrefmark{1},
Cheng Tan\IEEEauthorrefmark{2}\IEEEauthorrefmark{3},
Zeyu Li\IEEEauthorrefmark{1},
Jinming Ge\IEEEauthorrefmark{1},
Jiawei Liang\IEEEauthorrefmark{1},
Hao Yang\IEEEauthorrefmark{4},\\
Linfeng Du\IEEEauthorrefmark{1},
Jiang Xu\IEEEauthorrefmark{5},
Wei Zhang\IEEEauthorrefmark{1}\textsuperscript{\Letter}\thanks{\Letter\, Corresponding author: Wei Zhang (wei.zhang@ust.hk).}
}
\IEEEauthorblockA{
\IEEEauthorrefmark{1}The Hong Kong University of Science and Technology,
Hong Kong SAR, China\\
\IEEEauthorrefmark{2}Google, Mountain View, USA
\IEEEauthorrefmark{3}Arizona State University, Tempe, USA\\
\IEEEauthorrefmark{4}The George Washington University,
Washington, D.C., USA\\
\IEEEauthorrefmark{5}The Hong Kong University of Science and Technology (Guangzhou),
Guangzhou, China\\
\{shangkun.li, zliki, jgeab, jliangbr, linfeng.du\}@connect.ust.hk\\
chengtan@google.com, hao.yang@gwu.edu,
jiang.xu@hkust-gz.edu.cn, wei.zhang@ust.hk
}
}

\maketitle

\begin{abstract}
Coarse-Grained Reconfigurable Arrays (CGRAs) are a promising solution for achieving high energy efficiency and reconfigurability across various application domains, but their performance is often crippled by rigid memory architectures that limit the number and location of tiles that can access data memory. This creates a significant bottleneck for kernels with intensive memory accesses. To address this, we propose CMD, an integrated CGRA framework featuring cluster-based distributed memory design with a co-designed compilation toolchain. The compiler includes a novel memory-aware mapper and a design space exploration (DSE) mechanism that identifies the optimal memory architecture design for specific kernels. Experimental results show that our post-DSE CMD CGRAs achieve an average speedup of $1.39\times$ over a conventional CGRA while simultaneously reducing the total area to an average of $0.912\times$ of the conventional CGRA.
\end{abstract}

\begin{IEEEkeywords}
CGRA, Memory Architecture, Mapping.
\end{IEEEkeywords}

\section{Introduction}\label{sec:intro}
Coarse-Grained Reconfigurable Arrays (CGRAs) are accelerators consisting of an array of computing tiles interconnected via a Network-on-Chip (NoC). They strike a balance between high flexibility and energy efficiency \cite{2019SurveyonCGRA-ACMCompSurv}, making them prevalent for accelerating applications across domains from high-performance computing (HPC) \cite{2023RIKENCGRA-HEART} to machine learning (ML) \cite{2024SambaNova-HCS}. To accelerate kernels (typically performance-critical loop nests in an application), the compiler first transforms them into Data Flow Graphs (DFGs) \cite{2003LoopParaMap-DATE,2020OpenCGRA-ICCD, 2026ClusterCGRA-DATE} and then maps these DFGs onto the CGRA. The primary goal of this mapping is to minimize the initiation interval (II) --- the number of cycles between the start of successive loop iterations --- as it directly impacts total execution latency.

\begin{figure}[t]
    \centering
    \includegraphics[width=1.0\linewidth]{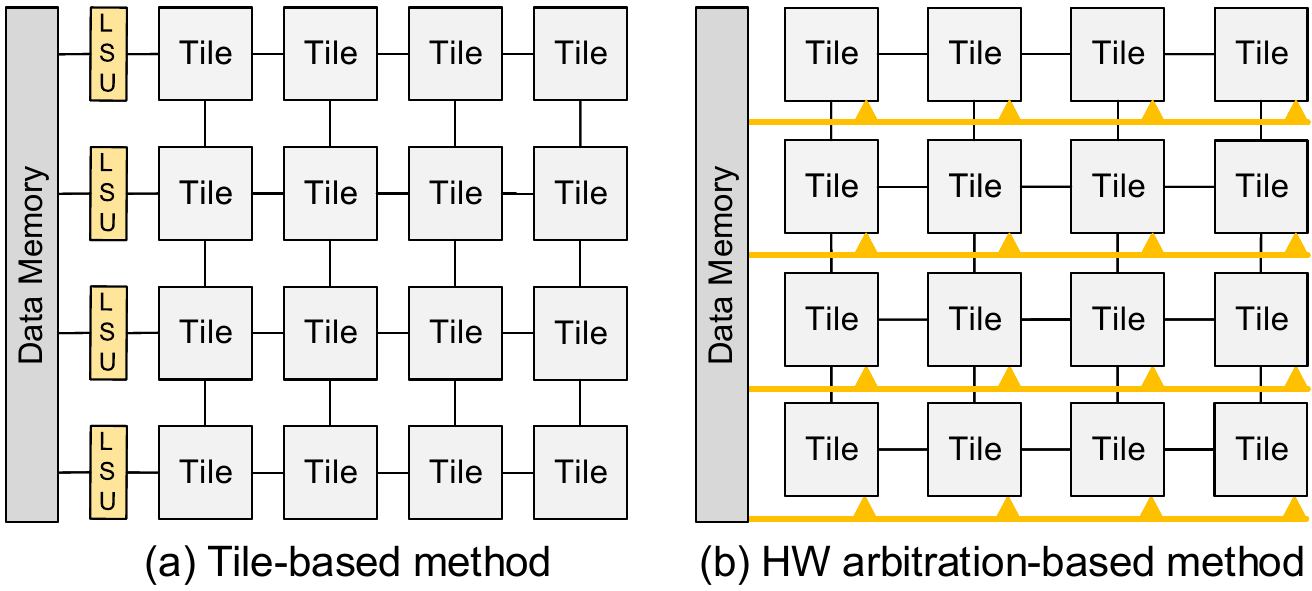}
    \vspace*{-1.4\baselineskip}
    \caption{Memory access methods --- (a) Tile-based method that enables the leftmost column to access the data memory. (b) HW arbitration-based method that enables one tile from each row to access the data memory in each cycle.}
    \label{fig:memory-access-method}
\end{figure}

However, the performance of CGRAs on kernels with memory access operations is often bottlenecked by the memory architecture. Most existing CGRAs use a global data memory adjacent to the tile array, allowing specific tiles to access memory each cycle. Two commonly used memory access methods are shown in Fig. \ref{fig:memory-access-method}: (1) The \textbf{tile-based method} \cite{2017HyCUBE-DAC,2023VecPAC-ICCAD} restricts memory access to a few fixed tiles --- typically those in the row or column closest to the data memory --- forcing data needed by other tiles to be relayed across the tile array. This creates severe routing pressure, which in turn inflates the II and degrades computation efficiency; (2) The \textbf{HW/SW arbitration-based method} \cite{2023Poly4CGRA-DAC,2022DRIPS-HPCA} either uses a hardware arbiter to select one tile per row or column to access the memory each cycle, or relies on the compiler to statically schedule non-conflicting memory accesses. While this method allows more tiles to potentially access memory, it still limits the number of tiles that can access memory each cycle. Specifically, when multiple tiles in the same row or column contend for memory access in the same cycle, this conflict forces some memory requests to be deferred or rerouted, increasing latency and data movement overhead. Fundamentally limited by a centralized memory system, both methods struggle to provide the parallel, low-latency data access required by many applications. Overcoming this bottleneck necessitates a paradigm shift to a distributed architecture with a co-designed compilation framework to fully exploit it.

In this paper, we take a significant step toward addressing these challenges by proposing CMD --- An Integrated \underline{C}GRA Framework with Cluster-Based Distributed \underline{M}emory \underline{D}esign. Our key contributions are summarized as follows:
\begin{itemize}
    \item \textbf{A Cluster-Based Distributed Memory Architecture} for CGRAs that alleviates memory access bottlenecks, featuring a lightweight controller to ensure coherence across distributed memory units. To the best of our knowledge, CMD is the first work to integrate distributed memory architecture design into a spatio-temporal CGRA; 
    \item \textbf{A Memory-Aware Kernel Mapper} that efficiently maps operations and variables of kernels onto the CMD CGRA;
    \item \textbf{A Memory Architecture DSE Mechanism} that automatically finds an optimal memory architecture design for a set of kernels from a domain or an application.
\end{itemize}

\section{Background and Motivation}\label{Bg&Mot}

\subsection{Mapping Constraints in CGRAs}\label{sec:constriants}
\begin{figure}[t]
    \centering
    \includegraphics[width=0.99\linewidth]{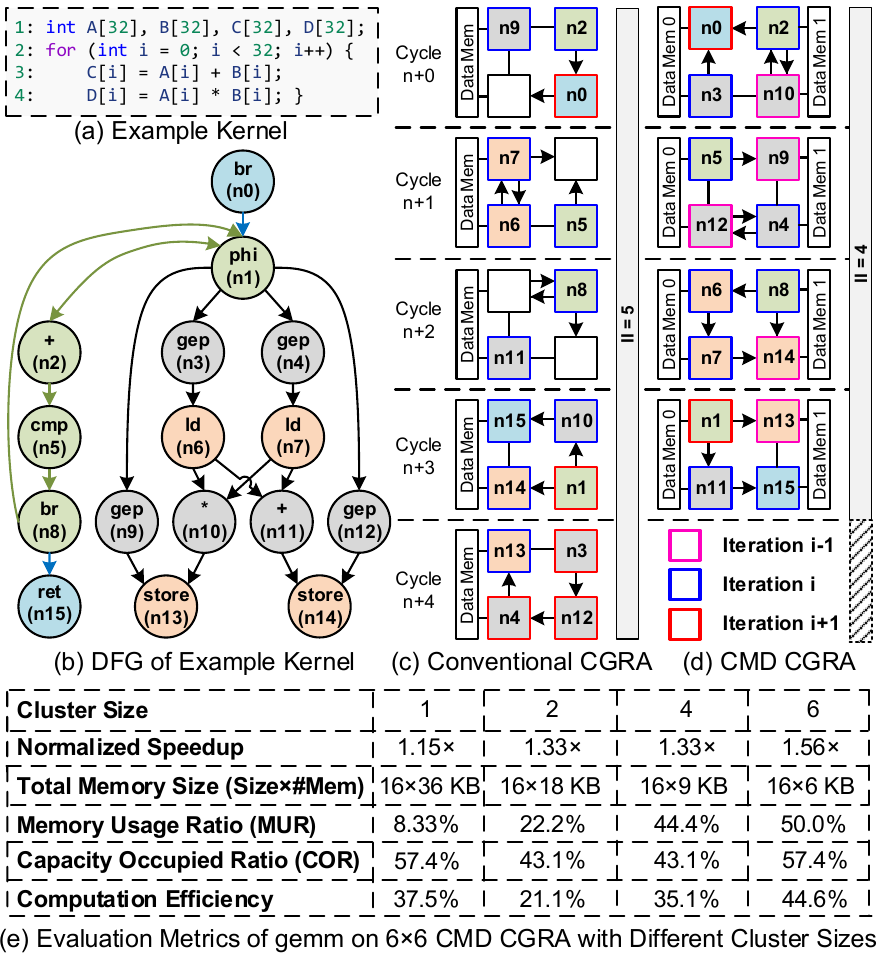}
    \vspace*{-1.4\baselineskip}
    \caption{Motivation Example --- (a) A vector operation kernel. (b) DFG of the kernel. Blue represents boundary-related operations. Green represents loop control operations. Grey denotes computational operations, where \texttt{gep} stands for the GetElementPtr \cite{2024LLVMGep-Online} operation. Orange indicates memory access operations. (c) Mapping the DFG onto a conventional CGRA. White tiles indicate tiles that are solely used for data transfer. (d) Mapping the EDFG (Extended-DFG, DFG with variable nodes) onto a CMD CGRA. (e) Mapping the EDFG of \texttt{gemm} onto $6\times 6$ CMD CGRAs with varying cluster sizes and the same memory unit size (16 KB).
    }
    \label{fig:motivation-example}
    \vspace*{-0.3\baselineskip}
\end{figure}

When using CGRAs to accelerate kernels, two main constraints heavily impact the mapping results: one arises from modulo scheduling \cite{1994ModuloScheduling-MICRO}, and the other stems from the limitations in conventional CGRA memory architectures. Accelerating applications on CGRAs involves mapping the DFGs onto the MRRG with the goal of minimizing the II, which means a new loop iteration starts every II cycles \cite{2002DRESC-FPT}. This leads to \textbf{inherent mapping constraints}, as the configuration of a CGRA repeats every II cycles. In a CGRA that uses the tile-based memory access method, as shown in Fig. \ref{fig:motivation-example}(c), only the tiles in the leftmost column are allowed to access memory. This limits both the number (i.e., two) and the positions (i.e., the leftmost column) of tiles that can access memory, thereby introducing the \textbf{routing constraints} during the spatio-temporal mapping \cite{2003LoopParaMap-DATE}. In architectures using the HW/SW arbitration-based memory access method, while each tile can theoretically access memory, the arbiter also restricts the number (i.e., one per row/column) and positions (determined by the HW/SW arbiter) of tiles that can access memory in each cycle. Thus, similar routing constraints still exist.

\subsection{Memory Bottlenecks in Conventional CGRA}\label{sec:bottleneck}

This work focuses on the \textbf{routing constraint}, which directly leads to two critical performance bottlenecks. We illustrate this by mapping a simple vector operation onto a $2\times2$ conventional CGRA with the tile-based memory access method (the leftmost column of tiles can access data memory) using a heuristic mapping algorithm \cite{2020OpenCGRA-ICCD}, as shown in Fig. \ref{fig:motivation-example}(a)-(c).



\textbf{Long II} --- The theoretical minimum II (MII) for this kernel is 4, determined by loop recurrence dependencies ($RecMII$) and computation resources ($ResMII$) \cite{2002DRESC-FPT}.
\vspace*{-0.4\baselineskip}
\begin{equation}\label{formula:mii}
\vspace*{-0.4\baselineskip}
    MII=max\{RecMII, ResMII\}
\end{equation}
However, the mapping result in Fig. \ref{fig:motivation-example}(c) shows an II of 5. This gap arises because the \textbf{routing constraint} is not captured by Eq. \eqref{formula:mii}. Ideally, if nodes \texttt{n13} and \texttt{n14} could be executed in the same cycle, we could complete all four memory access operations in two cycles and leave the rest of the tiles for other operations, thereby improving performance. Unfortunately, the \textbf{routing constraint} forces memory access operations (e.g., \texttt{n13}, \texttt{n14}) to queue up for the limited memory access resources, which stalls the computational operations that depend on them, preventing the mapper from fully utilizing the computational resources of the rest of the array (e.g., many tiles in Fig. \ref{fig:motivation-example}(c) are only used for relaying data). This inflated II directly increases the latency calculated by Eq. \eqref{formula2:latency}, where $TripCounts$ is the loop iteration count and $Steps$ is the execution cycles per iteration.
\vspace*{-0.3\baselineskip}
\begin{equation}\label{formula2:latency}
    Latency = II\times(TripCounts-1)+Steps
    \vspace*{-0.3\baselineskip}
\end{equation}

\textbf{Low Computation Efficiency} --- As shown in Fig. \ref{fig:motivation-example}(c), during cycles $n+0$ to $n+2$, several tiles are used only for data transfer rather than executing operations. To quantify this, we define \textbf{computation efficiency} as:
\begin{equation}\label{formula3:comp-efficiency}
    CompEff=\frac{1}{II}\sum_{i=0}^{II-1}\frac{N_{op}^i}{N_{op}^i+N_{tr}^i}
\end{equation}
As defined in Eq. \eqref{formula3:comp-efficiency}, where parameters are detailed in Table \ref{table:metrics}, a high number of transfer-only tiles (i.e., $N_{tr}^i$) reduces efficiency. This inefficiency is a direct consequence of the \textbf{routing constraint}. Since few tiles can access memory, other tiles are wasted on relaying data across the array, leaving fewer tiles for executing operations in each cycle, leading to poor parallelism and, consequently, higher latency. The problem is exacerbated in larger CGRAs (e.g., $6\times6$), where tiles farther from memory require more intermediate routing steps, further degrading computation efficiency.

\textbf{Benefits of CMD CGRA} --- Our proposed CMD CGRA directly addresses these bottlenecks. By organizing every two tiles into a cluster, each cluster has a memory unit accessible by the two tiles, which alleviates the \textbf{routing constraint} efficiently. As shown in Fig. \ref{fig:motivation-example}(d), this design allows mapping the Extended-DFG (EDFG, DFG with variable nodes, detailed in Sec. \ref{sec:DFG-construction}) onto the CMD CGRA with the optimal II of 4 and $100\%$ computation efficiency, yielding a $\mathbf{1.21\times}$ speedup over the conventional design in Fig. \ref{fig:motivation-example}(c).

\begin{table}[t]
\centering
\caption{Summary for Evaluation Metrics}
\vspace*{-0.7\baselineskip}
\resizebox{1.0\linewidth}{!}{ 
\begin{tabular}{c||c||c}
\toprule

\textbf{Metrics} & \textbf{Parameters} & \textbf{Descriptions} \\
\hline\hline
\multirow{2}{*}{\makecell[c]{$CompEff$}} & $N_{op}^i$ & \# of tiles for operations in cycle $i$ \\
    & $N_{tr}^i$& \# of tiles solely for data transfer in cycle $i$\\
\hline
\multirow{1}{*}{\makecell[c]{$C_{total}$}} & $C_i$ & Capacity of memory unit $i$ \\
\hline
\multirow{2}{*}{\makecell[c]{$MUR$}} & $N_{occupy}$ & \# of occupied memory units \\
    & $N$ & \# of all memory units \\
\hline
\multirow{1}{*}{\makecell[c]{$COR$}}& $C_{occupy}^i$ & Occupied size of memory unit $i$ \\
\bottomrule

\end{tabular}
}
\label{table:metrics}
\vspace*{-0.3\baselineskip}
\end{table}

\subsection{Design Space Exploration Opportunity}\label{sec:motivation-DSE}

\begin{figure*}[t]
    \centering
    \includegraphics[width=1.0\linewidth]{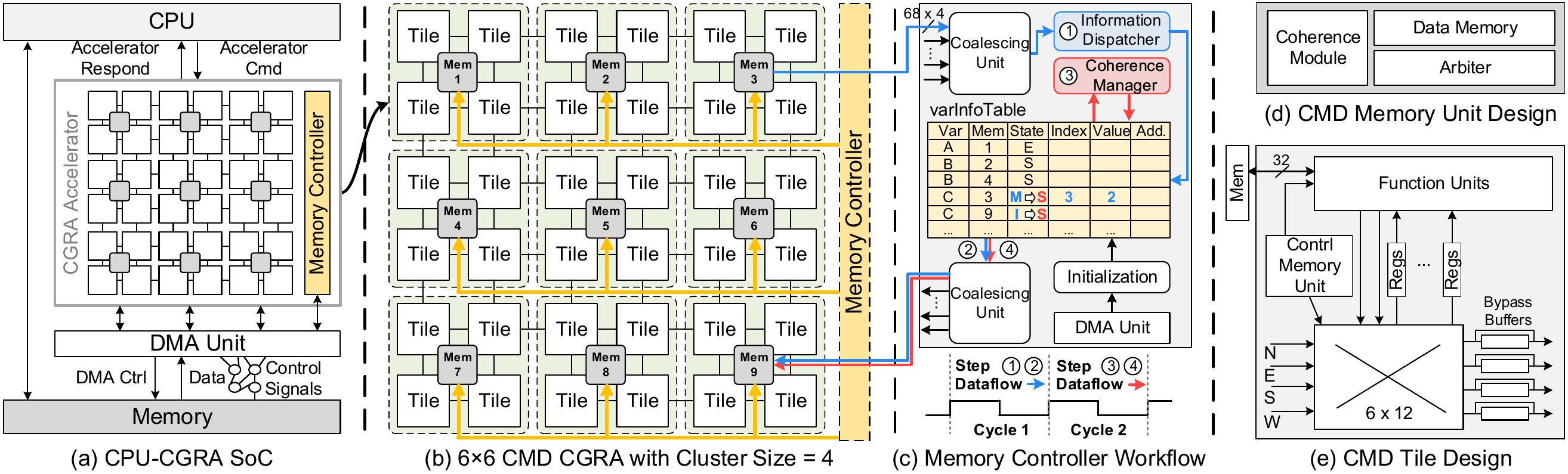}
    \vspace*{-1.2\baselineskip}
    \caption{CMD Architecture --- (a) The CMD CGRA is loosely coupled with the CPU, with the DMA Unit transferring data and control signals to the tiles, memory units, and memory controller. (b) In the CMD CGRA with a cluster size of 4, four tiles in each cluster share a memory unit, and all memory units connect to a memory controller. (c) The CMD memory controller collects signals generated by the coherence module in each memory unit and maintains a $varInfoTable$ to ensure memory coherence. (d) Each memory unit consists of a multi-bank data memory, a hardware arbiter, and a memory coherence module. (e) Each tile contains a function unit, a control memory, a $6\times 12$ crossbar, eight register sets, and four bypass buffers.}
    \label{fig:architecture}
    \vspace*{-1.2\baselineskip}
\end{figure*}

While powerful, the CMD CGRA's performance is sensitive to memory design parameters (i.e., the cluster size and memory unit size) due to diverse memory access patterns and data characteristics. We demonstrate this by mapping a \texttt{gemm} onto a $6\times6$ CMD CGRA with different cluster sizes. To demonstrate the impact of the cluster size, we fix the single memory unit size at 16 KB, as it --- not the total size --- directly influences the memory-aware mapping. To formally analyze the mapping results, we introduce three metrics (parameters detailed in Table \ref{table:metrics}).
The \textbf{Total Memory Size (\bm{$C_{total}$})} and \textbf{Memory Usage Ratio (MUR)} are defined as:
\par\nopagebreak
\noindent
\begin{minipage}{0.455\linewidth}
    \begin{equation}
        \label{formula3.5:totalMemSize}
        C_{total}=\sum_{i=1}^{N}C_i
    \end{equation}
\end{minipage}\hfill
\begin{minipage}{0.5\linewidth}
    \begin{equation}
        \label{formula4:memory-util}
        MUR=\frac{N_{occupy}}{N}
    \end{equation}
\end{minipage}
\par
A higher MUR indicates that more memory units are in use. We can use it to guide the selection of an appropriate cluster size for a given CGRA size. \textbf{Capacity Occupied Ratio (COR)} is defined as:
\begin{equation}\label{formula5:memory-eff}
    COR=\frac{\sum_{i=1}^N C_{occupy}^i}{\sum_{i=1}^N \alpha_i C_i},\ 
    \alpha_i=\left\{
\begin{aligned}
        &1,\ if\ Mem_i\ is\ occupied\\
        &0,\ otherwise
\end{aligned}
\right .
\end{equation}
COR evaluates the utilization of capacity in occupied memory units. It helps identify potential resource waste caused by mismatches between variable size and memory unit size.

Fig. \ref{fig:motivation-example}(e) shows that performance (e.g., Speedup) and hardware efficiency (e.g., MUR, COR, etc.) vary dramatically with different cluster sizes. A poorly chosen configuration (e.g., cluster size = 1) may achieve low speedup (i.e, $1.15\times$). Fig. \ref{fig:motivation-example}(e) also highlights variations in MUR, COR, and computation efficiency. Since both cluster size and memory unit size significantly impact the mapping results, a one-size-fits-all approach is insufficient. Therefore, to maximize performance while minimizing hardware overhead, a systematic DSE mechanism to co-optimize \textbf{cluster size} and \textbf{memory unit size} is essential for designing effective CMD CGRAs.

\section{CMD Architecture}\label{sec:architecture}
Fig. \ref{fig:architecture}(a) gives an overview of the CMD CGRA, operating as a loosely coupled accelerator controlled by a CPU. Fig. \ref{fig:architecture}(b) shows a $6\times6$ CMD CGRA, with tiles grouped into clusters of four tiles and each cluster sharing a local memory unit (detailed in Sec. \ref{sec:memory-architecture}). A memory coherence controller (detailed in Sec. \ref{sec:memory-controller}) maintains coherence across distributed memory units. Each tile contains function units that support LLVM \cite{2004LLVM-CGO} IR operations, a control memory, a $6\times 12$ crossbar, eight register sets, and four bypass buffers (Fig. \ref{fig:architecture}(e)). The tiles are interconnected in a mesh-style topology \cite{2009PPA-MICRO}.

\subsection{Distributed Memory Architecture}\label{sec:memory-architecture}
To address the \textbf{routing constraint} in conventional CGRAs, we propose a \textbf{cluster-based distributed memory architecture}. In this design, the CGRA is partitioned into clusters, where a group of tiles shares a dedicated memory unit, as shown in Fig. \ref{fig:architecture}(b). Each tile is restricted to accessing its local cluster's memory unit, subject to certain constraints (detailed in Sec. \ref{sec:mem-aware-mapping}). This significantly alleviates the constraints on the number and location of tiles that can access memory in conventional CGRAs. Our evaluation explores cluster sizes of $\{1,2,4,6\}$ for a $6\times6$ CMD CGRA. Other cluster sizes can also be supported based on user configurations.


\textbf{Memory Unit Design} --- As shown in Fig. \ref{fig:architecture}(d), the memory unit comprises a multi-bank data memory, with the number of banks matching the cluster size to enable simultaneous access from all tiles within the cluster, an arbiter for memory accesses, and a coherence module that works alongside the memory controller. 
We guarantee that kernels' variables fit into these memory units (e.g., using tiling). The coherence module monitors write operations. Upon detecting a write operation, it generates and sends a comprehensive signal packet (e.g., $68\times 4$ bits in Fig. \ref{fig:architecture}(c)) to the memory coherence controller. This packet contains the necessary information to maintain memory coherence, including which variable is modified, the specific index within the variable, the new value, etc. These memory units are loaded with data by the DMA Unit during configuration.


\subsection{Memory Controller}\label{sec:memory-controller}

Since a variable may be needed by tiles in multiple clusters, it may be stored in several memory units. This distribution necessitates a coherence mechanism to synchronize data after modifications. To address this, we introduce a memory coherence controller that implements a MESI-like protocol \cite{1984MESI-ISCA}. Since tile operations in every cycle are determined in the memory-aware mapping under certain rules (detailed in Sec. \ref{sec:mem-aware-mapping}), and combined with the proposed memory coherence controller, the overall memory consistency is guaranteed.


\textbf{Variable Information Table} --- As shown in Fig. \ref{fig:architecture}(c), the memory controller maintains memory coherence through a $varInfoTable$, which tracks every variable instance across the memory units. For each variable instance, the table records which memory unit it's in, its state ($modified$ (M), $exclusive$ (E), $shared$ (S), and $invalid$ (I)), the update details if modified (i.e., the index of the written data and the new value), and the head address of the variable in the memory unit. For example, as shown in Fig. \ref{fig:architecture}(c), writing to variable \texttt{C} in memory unit 3 changes its state to $Modified$ (M) while its other shared copy in memory unit 9 becomes $Invalid$ (I), with the table logging the precise index and value of the change. This table is initialized by the DMA Unit during configuration.

\textbf{Memory Coherence Mechanism} --- As a single-cycle implementation would degrade system frequency, while a longer delay would harm performance, we design a four-step memory controller that completes in two cycles to strike a balance, as shown in Fig. \ref{fig:architecture}(c). The process to handle a write operation, such as changing \texttt{C[3]} in memory unit 3 to 2, unfolds as follows. In Cycle 1, the information dispatcher interprets the input signal packet and updates the $varInfoTable$ -- setting \texttt{C}'s state in unit 3 to $Modified$ (M) and its shared copy in unit 9 to Invalid (I), recording the index and new value (Step \circled{\footnotesize{1}}) -- and then broadcasts an invalidation signal to unit 9 to invalidate unsynchronized variables (Step \circled{\footnotesize{2}}). In Cycle 2, the coherence manager identifies the $Invalid$ instances, generates a synchronization request with the new value, and updates the table state for both copies to $Shared$ (S) (Step \circled{\footnotesize{3}}). Finally, it sends this request to unit 9 to update its local data (Step \circled{\footnotesize{4}}).

\section{CMD Compilation Toolchain}\label{sec:compilation-framework}
\begin{figure*}[t]
    \centering
    \includegraphics[width=1.0\linewidth]{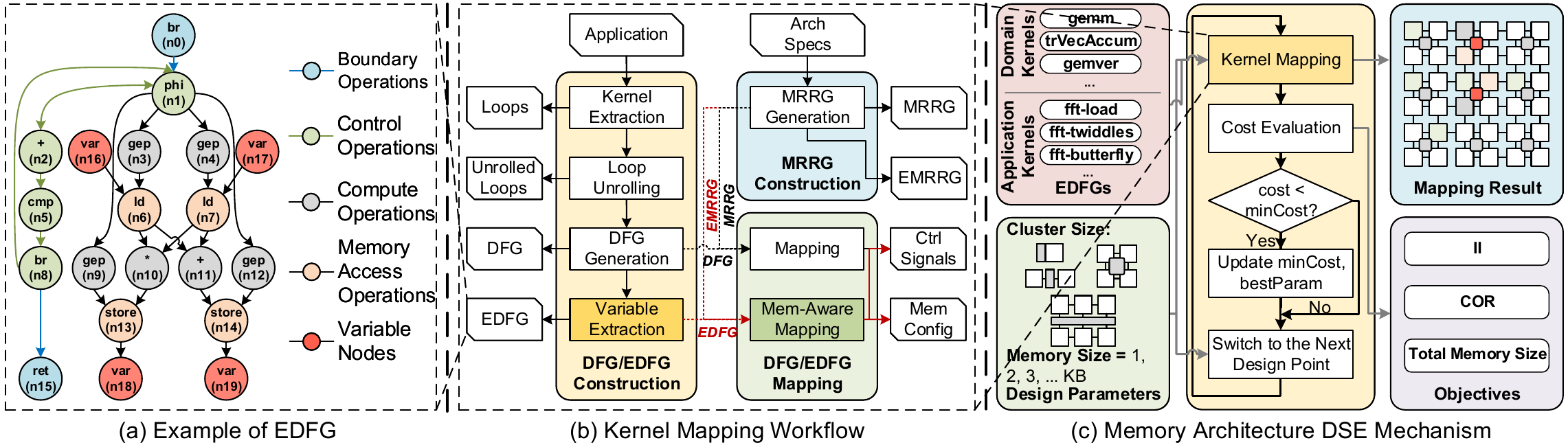}
    \vspace*{-1.2\baselineskip}
    \caption{CMD Compilation Toolchain --- (a) Each memory access operation (colored in orange) is accompanied by a variable node (colored in red) that records information about the variable in EDFG. (b) The kernel mapping supports mapping DFG onto MRRG (the black dotted line) and mapping EDFG onto EMRRG (the red dotted line) using memory-aware mapping. (c) The memory architecture DSE mechanism searches for the optimal design by analyzing the mapping results.
    }
    \label{fig:compilation}
    \vspace*{-1.0\baselineskip}
\end{figure*}
The CMD compilation toolchain consists of two components: kernel mapping and memory architecture DSE. As shown in Fig. \ref{fig:compilation}(b), the mapping flow first constructs a DFG/EDFG and generates the MRRG/Extended-MRRG (EMRRG, adding memory architecture information to the MRRG) based on the input architecture specifications. It then maps the DFG/EDFG onto the MRRG/EMRRG. Our mapping algorithm extends a heuristic mapping \cite{2020OpenCGRA-ICCD} with a proposed memory-aware mechanism. For conventional CGRAs, this mechanism is disabled, mapping the DFG onto the MRRG. For CMD CGRAs, it is enabled to map the EDFG onto the EMRRG. The memory architecture DSE mechanism resolves the trade-offs between performance and hardware overhead. It employs a grid search that evaluates multiple metrics to find an optimal architecture for a set of kernels from a specific domain or an application.


\subsection{DFG/EDFG Construction}\label{sec:DFG-construction}
\textbf{DFG Construction} --- A DFG represents kernels within an application, where nodes represent operations in LLVM \cite{2004LLVM-CGO} IR and edges indicate data dependencies between operations. Control dependencies are converted to data dependencies via partial predication \cite{2014Branch-DAC, 2026NEURA-PLDI} to be represented by the DFG.
Fig. \ref{fig:compilation}(b) shows that CMD extracts innermost loops from the application and constructs a DFG for them.


\textbf{EDFG Construction and Variable Extraction} --- To enable memory-aware mapping, we extend the DFG into an Extended-DFG (EDFG) by explicitly modeling variables. As shown in Fig. \ref{fig:compilation}(a), this EDFG corresponds to the kernel discussed in Sec. \ref{sec:bottleneck}. This EDFG creates variable nodes (\texttt{n16}, \texttt{n17}, \texttt{n18}, \texttt{n19}) and connects them to their corresponding memory access operation nodes (\texttt{n6}, \texttt{n7}, \texttt{n13}, \texttt{n14}). Each variable node (e.g., \texttt{n16}, representing vector \texttt{A[SIZE]}) encapsulates critical metadata, such as its size (e.g., 128 bytes), data type (e.g., \texttt{[32 x i32]}), variable name (e.g, \texttt{A}), associated memory access operation (e.g., \texttt{n6}), and a unique variable ID. For large-sized variables, the CMD compiler tiles the loop to partition them so that they can fit into memory units. Crucially, a variable accessed by multiple operations is initially represented by multiple distinct variable nodes, one for each access. This provides the memory-aware mapping with the flexibility to store the same variable across different memory units if beneficial for performance.

\subsection{Memory-Aware Mapping}\label{sec:mem-aware-mapping}
Our proposed memory-aware mapping algorithm (Algorithm \ref{alg:mem-aware-mapping}) maps the EDFG onto the EMRRG, aiming to minimize the II. The process begins by calculating a potential minimum II based on recurrence and resource constraints (line 1) \cite{2020OpenCGRA-ICCD}. It then enters an outer loop (line 2) that iteratively increases the II upon mapping failure. Within this loop, an EMRRG is generated for the current II (line 3), and the algorithm attempts to map each operation node (lines 4-16). For each operation node, it evaluates all available tiles (line 5), calculating a mapping cost using Depth-First Search for each valid tile (line 14). It then selects the optimal tile for placement and routing (lines 15-16). Two mechanisms are enforced during the mapping of memory access operations (lines 6-13):

\begin{algorithm}[t]
\footnotesize
\caption{CMD Memory-Aware Mapping}\label{alg:mem-aware-mapping}
\KwIn{EDFG, targetCGRA}
\KwResult{Control Signals, Memory Configuration}
II = InitializeMII(EDFG, targetCGRA); \\
\While{\upshape{No Available Mapping \textbf{and} II $<$ MaxII}}{
EMRRG = createEMRRG(targetCGRA, II);\\
\ForEach{\upshape{opNode $\in$ TopologicalOrder(EDFG)}}{
    \ForEach{\upshape{tile $\in$ Unmapped tiles of EMRRG}}{
        \If{\upshape{isMemAccess(opNode)}}{
        memoryUnit = getMemUnit(tile);\\
        checkAccessRules(varNode, opNode);\\
        \uIf{\upshape{var in memoryUnit}}{
        MergeVarNode(varNode);
        }{\ElseIf{\upshape{isOverflow(memoryUnit)}}{try next tile;}
        }
        varMap(varNode, memoryUnit);
        }
        tilesWithCost.insert(tile, calculateCost(opNode, tile));\\
    }
    optimalTile = min(tilesWithCost);\\
    placeAndRoute(opNode, optimalTile);\\
}
II = II + 1;
}
\end{algorithm}

\textbf{Variable Node Merging Mechanism} --- To improve data reuse and reduce memory waste, the algorithm employs a variable node merging mechanism. Since the EDFG may contain multiple variable nodes for the same variable, the algorithm checks if a variable already exists in the target memory unit before considering a placement (line 9). If so, the new variable node is merged with the existing one to avoid redundant storage (line 10). If not, it verifies that the memory unit has sufficient space to prevent overflow (lines 11-12).



\textbf{Rules for Memory Consistency} --- To guarantee overall memory consistency, two rules are enforced during mapping (line 8). (1) Since the memory coherence controller takes two cycles to complete, no other accesses to the same variable are allowed during this period after any modifications to the variable. This effectively prevents the use of stale variables. (2) When a variable is stored in multiple memory units, CMD restricts one write per cycle to the same index of the variable. This avoids simultaneous changes to the same data of the variable in different memory units. These mapping rules, together with the memory coherence mechanism, ensure \textbf{the overall memory consistency} of CMD CGRA.

\subsection{Memory Architecture DSE}\label{sec:architecture-refinement}
To find a robust architecture for specific kernels, the CMD compiler uses a grid-search-based memory architecture DSE mechanism, as shown in Fig. \ref{fig:compilation}(c). The input is a set of kernels from a \textbf{specific domain} or an \textbf{application}. This mechanism explores a design space defined by cluster size (e.g., $\{1,2,4,6\}$ for a $6\times 6$ CGRA) and memory unit size (1KB to a user-defined max). Since the design space is small, a grid search is feasible (completing in minutes). The cost of each candidate architecture is evaluated by averaging the costs of all input kernels. The goal is to find the architecture minimizing the overall cost. If no suitable design is found, we increase the CGRA size until a legal design is found.



\textbf{Cost Function} --- The DSE targets three objectives: minimizing II, maximizing COR, and minimizing Total Memory Size ($C_{total}$). MUR is not explicitly considered here, as it is highly correlated to $C_{total}$ and COR. Optimizing total memory size and COR inherently guides the MUR to a reasonable value. This multi-objective problem is formulated as a single cost function. The overall cost for a set of $M$ input kernels is the average of the costs for each kernel:
\begin{equation}\label{formula5:cost-function}
    cost=\sum_{k=1}^M(w_1\times II_k + w_2/COR_k +w_3\times C_{total,k})/M
\end{equation}
The weighted factors $w_i$ ($i=1,2,3$) allow for prioritizing design goals. In our experiment, we set $w_1\gg w_2\sim w_3$ to prioritize performance while also optimizing hardware overhead.

\section{Experiment Results}\label{sec:evaluation}

\begin{table*}[t]
\centering
\caption{Benchmarks for CMD Evaluation}
\vspace*{-0.7\baselineskip}
\resizebox{\linewidth}{!}{ 
\begin{tabular}{c||c||c|c||c|c|c|c||c|c}
\toprule

 \textbf{\multirow{2}{*}{Domains}} & \textbf{\multirow{2}{*}{\makecell[c]{Kernels/\\Applications}}} & \multicolumn{2}{c||}{\textbf{Variables}} & \multicolumn{4}{c||}{\textbf{EDFG}} & \multicolumn{2}{c}{\textbf{Perf/Area$^\mathrm{a}$}} \\
 \cline{3-10}
    &  & \#Vars& Total Size (bytes) & \#OpNds & \#Edges & \#VarNds & RecMII$^\mathrm{b}$ & Best Pre-DSE$^\mathrm{c}$ & Post-DSE \\
\hline\hline
\multirow{5}{*}{\makecell[c]{Linear\\Algebra}} & gemm & 3 & 21168 &33 & 48 & 4 & 4 & $1.56\times/1.20\times(\uparrow)$ & $1.55\times/0.937\times(\uparrow\uparrow)$ \\
    & 2mm & 5 & 35280 & 52 & 77 & 6 & 4 & $1.42\times/1.20\times$ & $1.63\times/0.937\times(\uparrow\uparrow\uparrow)$ \\
    & trVecAccum$^\mathrm{d}$ & 1 & 2048 & 18 & 28 & 1 & 4 & $1.40\times/1.20\times$ & $1.40\times/0.937\times(\uparrow\uparrow)$ \\
    & gemver & 5 & 6144 & 168 & 223 & 28 & 7 & $1.39\times/1.06\times(\uparrow)$ & $1.39\times/0.937\times(\uparrow\uparrow)$ \\
   & reduce-sum & 2 & 4352 & 73 & 121 & 8 & 5 & $1.30\times/1.20\times$ & $1.30\times/0.937\times(\uparrow)$ \\
\hline\hline
Signal Process & fft & 6 & 20480 & 96 & 134 & 24 & 7 & $1.26\times/1.20\times$ & $1.24\times/0.966\times(\uparrow)$ \\
Communication & viterbi & 4 & 940 & 97 & 152 & 8 & 8 & $1.49\times/1.20\times(\uparrow)$ & $1.49\times/0.746\times(\uparrow\uparrow\uparrow)$\\
 Graph  & floyd & 1 & 7056 & 20 & 30 & 4 & 4 & $1.21\times/1.20\times$ & $1.21\times/0.937\times(\uparrow)$ \\ 
 Optimization & levmarq & 2 & 84 & 52 & 71 & 4 & 3 & $1.40\times/1.20\times$ & $1.40\times/0.746\times(\uparrow\uparrow\uparrow)$ \\
ML & gcn & 11 & 55720 & 106 & 146 & 19 & 4 & $1.25\times/1.20\times$ & $1.25\times/1.04\times(\uparrow)$\\
\bottomrule

\multicolumn{10}{l}{$\mathrm{a}$. Perf/Area: $\uparrow$ for $1.2 \leq Perf/Area < 1.4$, $\uparrow\uparrow$ for $1.4 \leq Perf/Area < 1.7$, $\uparrow\uparrow\uparrow$ for $Perf/Area \geq 1.7$.}\\


\multicolumn{10}{l}{$\mathrm{b}$. The RecMII represents the minimum II determined by the inter-iteration data dependencies.}\\

\multicolumn{10}{l}{$\mathrm{c}$. Best Pre-DSE: the cluster size from \{1, 2, 4, 6\} that yields the highest speedup for that kernel.}\\

\multicolumn{10}{l}{$\mathrm{d}$. trVecAccum transforms a triangular matrix.}\\
\end{tabular}}
\vspace*{-1.2\baselineskip}
\label{table:benchmarks}
\end{table*}
The CMD compiler is built upon LLVM 10.0.0 \cite{2004LLVM-CGO}, integrating a cycle-accurate simulator for performance evaluation. We implement the CMD CGRA in RTL \cite{2020OpenCGRA-ICCD, 2023VecPAC-ICCAD} and synthesize it with Synopsys Design Compiler V-2023.12 and CACTI 7.0 \cite{2009CACTI-HPL} to get the power, area, and timing. We evaluate CMD using a diverse suite of benchmarks sourced from \cite{2012PolyBench-url, 2014MachSuite-ISWC, 2022DASS-TCAD, 2021AutoSA-FPGA, 2020OpenCGRA-ICCD}, which form two categories: a focused domain of linear algebra kernels to test our performance and domain-specific DSE, and a diverse set of complex applications from different domains (e.g., signal processing, graph, ML, etc.) to evaluate broader effectiveness, as shown in Table \ref{table:benchmarks}. The performance is compared against a baseline conventional $6\times6$ mesh CGRA \cite{2020OpenCGRA-ICCD} ($0.36\ \text{mm}^2$ with SRAM), which is configured with a 64 KB global SRAM -- a minimum size sufficient to execute all benchmarks. To ensure a fair comparison, the baseline mapper is identical to the CMD mapper but without our memory-aware mechanism.

\subsection{Performance and Memory Evaluation}\label{sec:experiment-evaluation}
\begin{figure}[t]
    \centering
    \includegraphics[width=1.0\linewidth]{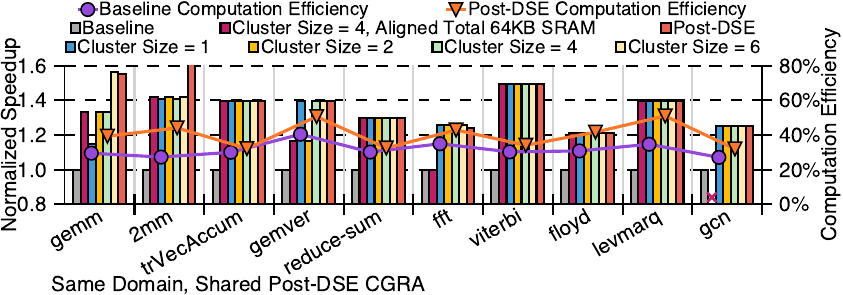}
    \vspace*{-1.85\baselineskip}
    \caption{Performance Evaluation -- Normalized speedup of pre-DSE CMD CGRAs (with 16 KB memory units and cluster sizes of \{1,2,4,6\}) and the post-DSE design, compared to the baseline. The figure also shows the computation efficiency of the post-DSE design compared to the baseline.}
    \label{fig:performance-evaluation}
\end{figure}

\textbf{Pre-DSE Evaluation} --- We begin by evaluating a $6\times6$ CMD CGRA with a cluster size of 4, aligning its total SRAM size with the baseline's 64KB. However, it fails to map the 
\texttt{gcn} due to insufficient per-cluster memory unit size, as shown in Fig. \ref{fig:performance-evaluation}. To enable a full-suite pre-DSE evaluation, we adopt a 16 KB memory unit per cluster --- the minimum size dictated by the most demanding kernel \texttt{gcn} --- for $6\times 6$ CMD CGRAs with cluster sizes of $\{1, 2, 4, 6\}$. As shown in Fig. \ref{fig:performance-evaluation}, CMD achieves respective average speedup of $\{1.33\times, 1.32\times, 1.34\times, 1.37\times\}$ for cluster sizes of $\{1, 2, 4, 6\}$ over the baseline, with an overall average speedup of $1.34\times$. Although they bring a larger area overhead, the performance gains yield superior area efficiency (performance per area), as detailed in Sec. \ref{sec:physical-design}.

\textbf{Impact of Rules for Memory Consistency} --- We conduct an ablation study by removing rules introduced in Sec. \ref{sec:mem-aware-mapping} from our mapper, which results in an \textbf{identical latency} for our benchmarks. This is because the latency of rule (1) is hidden by mapping other independent operations in these cycles. The write restriction in rule (2) is used to ensure correctness and does not impact performance, as such write conflicts should not be present in a valid kernel.

\begin{figure}[t]
    \centering
    \includegraphics[width=1.0\linewidth]{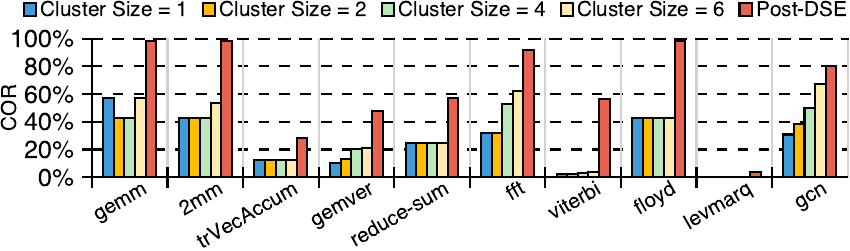}
    \vspace*{-1.7\baselineskip}
    \caption{Capacity Occupied Ratio --- COR (defined in Eq. \ref{formula5:memory-eff}) of cluster sizes being $\{1,2,4,6\}$ with 16 KB memory unit and the post-DSE architecture.}
    \label{fig:memory-eff}
    \vspace*{-0.5\baselineskip}
\end{figure}

\textbf{Post-DSE Evaluation} --- We apply the memory architecture DSE mechanism to find the optimal cluster size and memory unit size via two strategies: (1) a \textbf{domain-specific} search to find an optimal design for kernels in the linear algebra domain, and (2) a \textbf{per-application} search to find the optimal design for kernels in each application. As shown in Fig. \ref{fig:performance-evaluation}, the post-DSE designs achieve an average speedup of $1.39\times$ (up to $1.63\times$) over the baseline. The DSE mechanism intelligently balances performance against hardware overhead under the selected cost function's weighted factors. For example, the post-DSE design for \texttt{gemm} accepts a minor speedup reduction (compared with a cluster size of 6 with 16 KB memory unit) to gain a significant improvement in COR, as shown in Fig. \ref{fig:memory-eff}. Moreover, the post-DSE architectures achieve an average computation efficiency of $40.0\%$ (up to $51.0\%$). This $1.27\times$ improvement over the baseline validates that CMD CGRA's flexible memory access frees more tiles for computation rather than mere data routing.

\begin{figure}[t]
    \centering
    \includegraphics[width=1.0\linewidth]{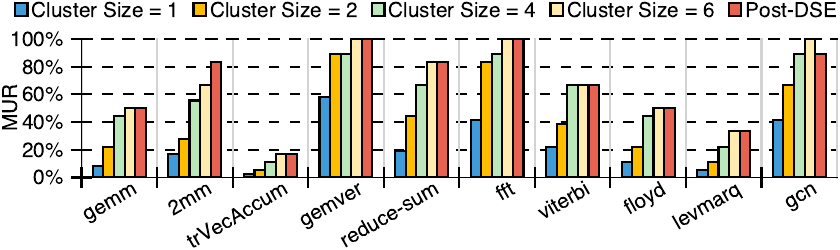}
    \vspace*{-1.7\baselineskip}
    \caption{Memory Usage Ratio --- MUR (defined in Eq. \ref{formula4:memory-util}) of cluster sizes being $\{1,2,4,6\}$ with 16 KB memory unit and the post-DSE architecture.}
    \label{fig:memory-util}
\end{figure}

\begin{figure}[t]
    \centering
    \includegraphics[width=1.0\linewidth]{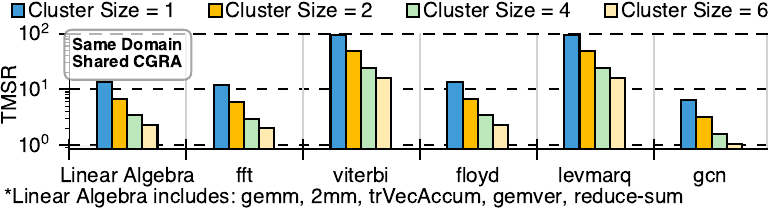}
    \vspace*{-1.4\baselineskip}
    \caption{Total Memory Size Reduction (TMSR) --- The factor by which the $C_{total}$ (defined in Eq. (\ref{formula3.5:totalMemSize})) of the post-DSE architecture is reduced compared to the architectures with 16 KB memory unit and cluster sizes of \{1,2,4,6\}.}
    \label{fig:memory-size}
\end{figure}

\textbf{Memory Analysis for Post-DSE Architecture} --- A key benefit of our DSE is the substantial reduction of memory hardware overhead, quantified by MUR, COR, and Total Memory Size Reduction (TMSR). Fig. \ref{fig:memory-util} shows that CMD can find a suitable cluster size to reduce the number of idle memory units and achieve an average MUR of $67.2\%$ (up to $100\%$), resulting in an overall average improvement of $1.70\times$ over the pre-DSE architectures. Memory unit size is also adjusted to an appropriate size to avoid memory waste, leading to an average COR of $66.1\%$ (up to $98.2\%$), achieving an average $2.34\times$ improvement over the pre-DSE architectures. Furthermore, the DSE dramatically reduces the total memory size. As shown in Fig. \ref{fig:memory-size}, the post-DSE architectures slash the total memory size by an average of $13.9\times$ over the pre-DSE architectures.


\textbf{Discussion on DSE} --- In our design space, some design points are unreachable due to interdependencies among evaluation metrics. Assigning weighted factors to each objective in the cost function further constrains the design space, inherently introducing trade-offs that prevent the simultaneous optimization of all objectives. Instead, we focus on minimizing the overall cost based on design requirements. In our experiment, the chosen weighted factors prioritize a low II while penalizing excessive memory overhead. This allows the DSE to intelligently trade performance for area savings, or vice versa, to identify the most cost-effective design.


For instance, some post-DSE architectures may exhibit a lower MUR than a pre-DSE design (e.g., MUR for \texttt{gcn} in the post-DSE architecture is lower than the pre-DSE architecture with a cluster size of 4). This is because the DSE mechanism finds that a substantial improvement in a heavily-weighted metric (e.g., COR in this case) justifies a minor sacrifice in other metrics (e.g., MUR in this case) to achieve a better overall cost function. This demonstrates that our DSE mechanism effectively balances competing objectives through the chosen weighted factors, tunable to meet different design goals.

\subsection{Timing, Area, and Power}\label{sec:physical-design}
We synthesize a prototype $6\times6$ mesh-style CMD CGRA (cluster size 4, 16 KB memory unit per cluster) using the TSMC
22nm ULL library at 800MHz. The entire area with SRAM is $0.38\ \text{mm}^2$ (Tiles $0.25\ \text{mm}^2$, Controller $0.0019\ \text{mm}^2$, SRAM $0.13\ \text{mm}^2$), with an average power of 228.69 mW (Tiles 163.63 mW, Controller 0.36 mW, SRAM 64.70 mW). The memory controller accounts for only $0.50\%$ of the area.


We evaluate the normalized performance per area (Perf/Area, with SRAM area) of the CMD CGRA against the baseline. As shown in Table \ref{table:benchmarks}, our best pre-DSE configurations ($6\times6$ with cluster size of 4 for \texttt{gemver} and 6 for the rest) achieve speedups that consistently surpass their area overhead of $1.20\times$ ($1.06\times$ for \texttt{gemver}). The benefits are further amplified post-DSE: when tailored for kernels from specific domains/applications, CMD achieves an average $1.39\times$ speedup over the baseline while reducing the area to an average of $0.912\times$, demonstrating high area efficiency.

To validate the scalability of the memory controller, we synthesize the memory controller for various CGRA sizes (Table \ref{table:physical-evaluation}). While the controller's absolute area grows with the array size, its \textbf{Ctrl. Overhead} (defined as the ratio of controller area to tiles area) steadily decreases. This trend confirms our controller is highly scalable and does not introduce an area bottleneck in larger CGRAs.



\begin{table}[t]
\centering
\caption{Scalability and Area of the Memory Controller}
\vspace*{-0.7\baselineskip}
\resizebox{1.0\columnwidth}{!}{ 
\begin{tabular}{c||c||c||c||c||c}
\toprule

\textbf{CGRA Size$^\mathrm{a}$} & $\bm{6\times 6,\,4}$ & $\bm{6\times 6,\,6}$ & $\bm{8\times 8,\,4}$ & $\bm{16\times 16,\,4}$ & $\bm{16\times 16,\,8}$\\
\hline\hline
Tiles Area (mm$^2$) & $0.254$ & $0.254$ & $0.451$ & $1.80$ & $1.80$ \\
Controller Area (mm$^2$) & 0.00187 & 0.00130 & 0.00297 & 0.00429 & 0.00309\\
Ctrl. Overhead$^\mathrm{b}$ & 0.00736 & 0.00512 & 0.00659 & 0.00238 & 0.00172\\
\bottomrule

\multicolumn{6}{l}{$\mathrm{a}$. CGRA Size stands for \{array size, cluster size\}.
}\\

\multicolumn{6}{l}{$\mathrm{b}$. Ctrl. Overhead is the controller area normalized to the total area of all tiles.}
\end{tabular}
}
\vspace*{-0.45\baselineskip}
\label{table:physical-evaluation}
\end{table}

\section{Related Works}\label{related works}
\textbf{CGRA Memory Architectures} -- Most CGRAs place data memory adjacent to the tile array, accessing it via the tile-based method \cite{2017HyCUBE-DAC, 2023VecPAC-ICCAD, 2023RIKENCGRA-HEART, 2022RipTide-MICRO} or the arbitration-based method \cite{2003ADRES-FPL, 2022DRIPS-HPCA, 2009PPA-MICRO}. These architectures suffer from the memory bottlenecks discussed in Sec. \ref{sec:bottleneck}. While Plasticine \cite{2017Plasticine-ISCA} and Xilinx AIE \cite{2025ESFA-TCAD} use distributed memory, their designs serve different purposes. Plasticine is designed for specific parallel patterns, and AIE's memory-per-tile design limits data reuse. Moreover, neither is designed for spatio-temporal CGRAs, nor do they provide a DSE framework to co-optimize the memory architecture, which CMD directly addresses.

\textbf{Memory-Aware Kernel Mapping} -- Existing compiler efforts focus on optimizing data access for conventional CGRAs. These include introducing cost functions in mapping for data reuse \cite{2011MemMap-TODAES}, replacing long data dependencies with memory accesses \cite{2016MemMap-VLSI, 2018RAMP-DAC}, or grouping memory access operations that access the same cache line into the same partition of DFG to increase cache line reuse \cite{2022ReuseAwarePartition-ISPA}. \cite{2023Poly4CGRA-DAC} presents a polyhedral-based loop transformation to improve data reuse for imperfectly nested loops. Morpher \cite{2022Morpher-WOSAET} can extract data layouts and distribute data uniformly in memory, but it does not consider the mapping of memory access operations. However, these works are tailored for conventional, centralized memory systems and operate on standard DFGs. They fundamentally lack the mechanism to manage the complex data placement and variable merging required by a distributed memory architecture. CMD addresses this by introducing EDFG and a novel memory-aware mapping.

\section{Conclusion}
This paper proposes CMD, an integrated framework with a cluster-based distributed memory architecture and a co-designed compiler. This compiler features a memory-aware mapping algorithm and a memory architecture DSE mechanism to find an optimal design that balances performance and hardware overhead for specific kernels. Evaluations show that the post-DSE CMD CGRAs achieve an average speedup of $1.39\times$ over baseline, while using only an average of $0.912\times$ the total area of the baseline.

\section*{Acknowledgments}
 This work was supported by the Research Grants Council (RGC) of Hong Kong SAR under the General Research Fund (GRF) \#16218324 and the Strategic Topics Grant (STG) \#STG3/E-605/25-N. We would also like to thank the anonymous reviewers for their insightful comments.

\bibliographystyle{IEEEtran}
\bibliography{refs}

\end{document}